\documentclass[epj]{svjour}
\usepackage[T1]{fontenc}
\usepackage[latin1]{inputenc}
\usepackage{graphics}
\usepackage{amsmath}
\usepackage{amstext}
\usepackage{epsfig}
\usepackage{epic}
\usepackage{eepic}
\usepackage{amssymb}
\usepackage{exscale}
\usepackage{pst-node,array}

\begin{document}
\title{Parity law of the singlet-triplet gap in graphitic ribbons}
\author{Mohamad Al $\mbox{Hajj}^{(a)}$, Fabien $\mbox{Alet}^{(b)}$, Sylvain $\mbox{Capponi}^{(b)}$, Marie Bernadette 
$\mbox{Lepetit}^{(c)}$, Jean-Paul $\mbox{Malrieu}^{(a)}$ and Synge $\mbox{Todo}^{(d)}$}
\institute{$(a)$ Laboratoire de Chimie et Physique Quantique, IRSAMC/UMR5626, Universit\'e Paul Sabatier, 118 route
de Narbonne, F-31062 Toulouse Cedex 4, France \\
$(b)$ Laboratoire de Physique Théorique, IRSAMC/UMR5152, Universit\'e Paul Sabatier, 118 route
de Narbonne, F-31062 Toulouse Cedex 4, France \\ 
$(c)$ CRISMAT, 6 Bd Mar\'echal Juin, F-14050 Caen Cedex, France \\
$(d)$ Department of Applied Physics, University of Tokyo, Tokyo 113-8656, Japan.}
\authorrunning{M. Al Hajj et al.}
\titlerunning{Parity law of the singlet-triplet gap}
\date{}
\abstract{
This work explores the possibility to transfer the parity law of the singlet-triplet gap established for square ladders 
(gapped for even number of legs, gapless for odd number of legs) to fused polyacenic 1-D systems, i.e., graphite ribbons. 
Qualitative arguments are presented in favor of a gapped character when the
number $n_{\omega}$ of benzene rings along the ribbon width is odd. 
A series of numerical calculations (quantitative mapping on spin $1/2$ chains, renormalized excitonic treatments and 
Quantum Monte Carlo) confirm the parity law and the gapless character of the ribbon for even $n_{\omega}$.
}
\maketitle
\section{Introduction}
In the recent past the properties of some quasi 1-D strongly correlated materials, namely cuprate ladders, have attracted 
much interest from solid state physicists~\cite{Dagotto}. As a major result they have established that the spin gap (i.e., lowest singlet 
to triplet excitation energy) of spin $1/2$ ladders presents a parity law: the ladders are gapped (have a finite excitation energy) 
when the number of legs is even, and gapless (degenerate singlet and triplet state) for odd number of legs \cite{Ref1}.
This result was not expected, since one may consider the ladders as intermediate between the simple 1-D chain and the square 
2-D lattice, which are both gapless.

Finite ribbons of fused polyacenes 
(with CH bonds on the most external doubly-bonded carbons) can be seen as organic analogs of the cuprate 
ladders. They are quasi 1-D fragments of graphite, which is gapless. The non-dimerized linear polyene  
is also known to be gapless. In light of the recent investigations on both
carbon and spin nanotubes, we can extend the discussion on the link between magnetic
properties and geometry of such objects~\cite{Matsumoto}. In particular, are the fused polybenzeno\"id ribbons always gapless ?

Although the $\pi$-electrons of the conjugated hydrocarbons are not strongly correlated, it has been shown twenty years 
ago that they can be treated accurately through $S=1/2$ Heisenberg Hamiltonians 
\begin{equation}
\label{eq:Heis}
H = J \sum_{\left< i,j \right>} {\bf S}_i \cdot {\bf S}_j
\end{equation}
{\it i.e.}, as spins interacting with their nearest neighbours through
antiferromagnetic (AF) couplings $J>0$~\cite{Ref2,Ref3}. This effective model is not based on the usual perturbative approach in the strong coupling limit;
instead the amplitude
of the exchange integral is given by the exact solution of the 2-site
problem. More explicitely, for a Hubbard Hamiltonian, the AF coupling is no longer taken as its
perturbative estimate $J=4 t^2/U$, but rather as the exact energy difference
between the triplet and singlet states $J=- (U-\sqrt{U^2+16t^2})/2.$
This view of $\pi$-electron systems offers simple rationalizations of many of their 
properties. In particular, a geometry-dependent Heisenberg Hamiltonian
(with $J(r)$ depending on the distance $r$ between sites) has
been extracted from accurate calculations of the singlet and triplet states of ethylene and happens to be a quantitative tool for the ground and lowest 
excited states of conjugated hydrocarbons \cite{Ref2,Ref3} (a
spin-independent $V(r)$ potential is of course also needed in that case to take into
account the $r$-dependences of the localized $\sigma$-bonds). This analysis
was proved in various situations to be very predictive: for instance, the ground-state geometries of a large series of hydrocarbones are accurately
reproduced by minimizing the lowest eigenvalue of the geometry-dependent
Hamiltonian with respect to the bond distance \cite{Ref2}. It was also shown
that the vertical excitation spectrum to the lowest triplet states is in good agreement with
experiments \cite{Ref3} and that the excited state geometries and vertical
emission energies \cite{Ref3} agree with accurate {\it ab initio}
calculations. This magnetic model has been widely and successfully used by
Robb {\it et al.} \cite{Ref4} under the acronym MM-VB (Molecular Mechanics
Valence Bond) for the study of the  photochemistry of conjugated hydrocarbons.

\begin{figure}[t]
\centerline{\includegraphics[scale=1.0]{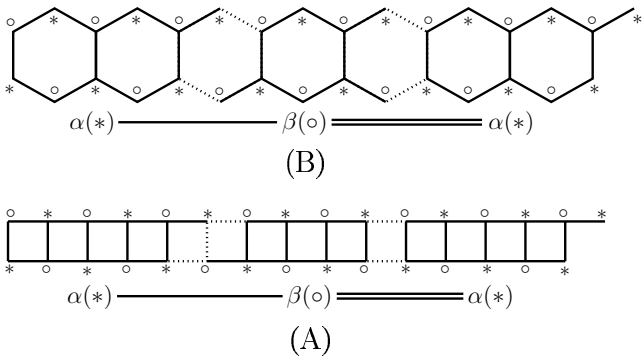}}
\caption{Two-leg ladder (A) and $n_{\omega}=1$ fused polyacene (B), and their mapping into dimerized spin 1/2 chains. 
In this figure and the following ones, the symbols $\circ$ and $*$ correspond to up and down spins respectively;
the resulting effective models appear just below each lattice. We denote with a single- and double-line the effective couplings when they are different.
For the graphitic ribbons, dotted lines separate the different blocks used in the calculation (see text). 
}
\label{fig1}
\end{figure}

In fact, it is quite natural to expect, at least for bipartite lattices, the
lowest excitation energies to be correctly described by a (properly
normalized) Heisenberg Hamiltonian as the lowest states are dominated by
neutral VB configurations and as the antisymmetrization favours an
antiferromagnetic order independently of the value of
$U$~\cite{Lepetit89}. Of course, the charge excitations, essentially  leading to
ionic states, are not accessible with these approaches, but it is
known for idealized 1D chains~\cite{lieb} and from experiments on conjugated
molecules~\cite{hudson} that the dipolarly-allowed VB ionic states are located at higher
energies than the lowest triplet (and even than neutral singlet states).

We now only consider the purely magnetic model of Eq.~(\ref{eq:Heis}). 
A few years ago density matrix renormalization group (DMRG)
calculations have been reported, concerning the singlet-triplet gap of 
the simplest polyacenic chain, built of aligned fused benzene rings \cite{Ref5}. The extrapolated calculated gap is 
finite and close to $0.1J$. Actually the polyacene can be viewed as 
a two-leg ladder in which one rung over two has vanished (see Fig.~\ref{fig1}). One might wonder whether there is a 
similarity between the three-leg ladder and a fused polyacenic infinite
ribbon with two ranks of benzene rings (see Fig.~\ref{fig3}).
From qualitative arguments one may 
conjecture that the singlet-triplet excitation in such polyacenic ribbons of graphite is finite when the 
number of superposed rings in the width of the ribbon, $n_{\omega}$, is 
odd and vanishes when $n_{\omega}$ is even. Numerical calculations using the quantitative mapping on known 1-D chains, 
renormalized excitonic method (REM) \cite{Ref5bis}, and Quantum Monte Carlo calculations (QMC), show that the gap
indeed vanishes for even $n_{\omega}$ and is non-zero for odd $n_{\omega}$ ribbons.
\section{Qualitative arguments}
\begin{figure}[b]
\centerline{\includegraphics[scale=1.0]{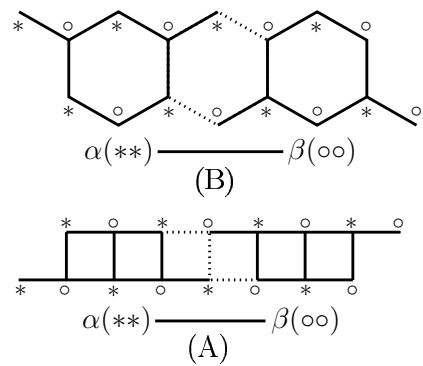}}
\caption{Mapping of ladder and polyacene into a spin-1 chain.}
\label{fig2}
\end{figure}

Qualitative arguments can be used to rationalize the parity law of ladders, which can also be applied to 1-D fused 
polyacenes. They are based on the real-space renormalization group (RSRG), originally proposed by Wilson \cite{Ref6}. 
Concerning spin 
lattices one may consider that they are built from blocks rather than from sites. Of course these blocks interact. 
If the blocks do not have a singlet ground state, but a doublet, or a triplet, they can be seen as interacting effective 
spins \cite{Ref7}. A quasi 1-D lattice can then be easily transformed \cite{Ref8} into a simple 1-D spin chain, 
the properties of 
which are well known. Considering 2-leg ladders, one may define $(2N_s+1)$ sites blocks, which have a ground state 
doublet $(S_z=\pm1/2)$. But these blocks do not have equal interactions with their left and right nearest neighbors 
(cf. figure \ref{fig1} (A)). Consequently, the 2-leg ladder maps into a
\textit{dimerized} (i.e., bond-alternated) spin $1/2$ chain, which 
is known to be gapped. An alternative partition uses blocks with even number
of sites (figure \ref{fig2} (A)) which have a triplet ground-state. The
resulting effective $S=1$ chain exhibits the famous Haldane gap \cite{Ref9} and therefore both partitions predict a gap. 

If one applies the same arguments to the $n_{\omega}=1$ polyacenic
chain one obtains a similar mapping pictured in figure \ref{fig1} (B). The partition into blocks of 
$(4N_S+1)$ sites and $(4N_S+3)$ sites produces an alternating dimerized spin chain, which is gapped. The partition 
into $4N_S$ sites blocks with triplet ground states (figure \ref{fig2}) leads to a Haldane gap \cite{Ref9}. Both 
partitions suggest a finite excitation energy. 

\begin{figure}[t]
\centerline{\includegraphics[scale=1.0]{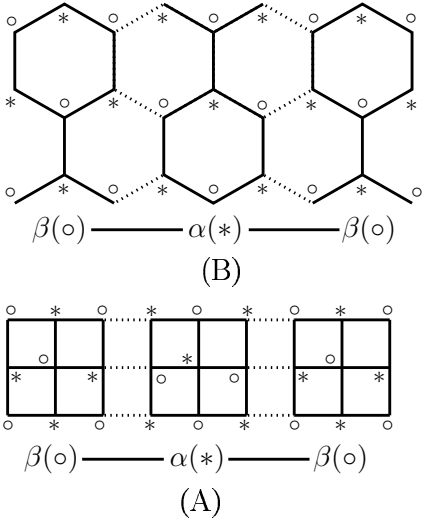}}
\caption{Three-leg ladder (A) and $n_{\omega}=2$ fused polyacene (B), and their mapping into spin 1/2 chains.}
\label{fig3}
\end{figure}

The ladder with odd number of legs can be partitioned into blocks with 
odd number of sites which have equal interactions with the left and right
neighbors, and the ladders can be mapped into a 
non-dimerized gapless 1-D $S=1/2$ chain, (cf. figure \ref{fig3} (A)). For the $n_{\omega}=2$ polybenzeno\"id ribbon the simplest 
partition defines 9-site blocks presenting a $S=1/2$ ground state and equal AF interactions 
with left and right nearest neighbors (cf. Fig. \ref{fig3} (B)). The
resulting mapping to an effective AF uniform chain indicates that the $n_{\omega}=2$ polyacenic ribbon should 
not be gapped. This is actually one of the two possibilities allowed by the famous Lieb-Schultz-Mattis theorem~\cite{LSM}
which can be applied for even $n_\omega$ (where the unit cell contains an odd number of spin $1/2$). 
If even  $n_\omega$ lattices do have a gap, this theorem guarantees that the system breaks translation symmetry.

\begin{figure}[h]
\centerline{\includegraphics[scale=1.0]{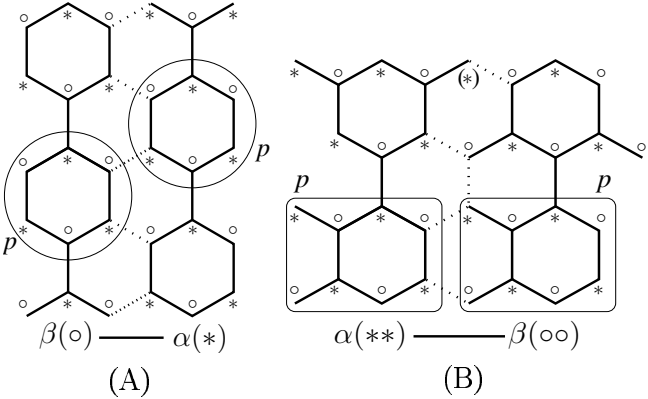}}
\caption{Generalization of the mappings on non-dimerized AF $S=1/2$ chains for 
even $n_{\omega}$ (A) and $S=1$ chains for odd $n_{\omega}$ (B). Moving the $(\ast)$ atom to the right-side block 
leads to an effective dimerized $S=1/2$ AF chains.}
\label{fig4}
\end{figure}

The arguments can easily be generalized to any thickness of the ribbon according to figure \ref{fig4}, 
which proposes a mapping 
into an uniform $S=1/2$ spin chain for even values of $n_{\omega}$ and into
an uniform integer-spin $S$ chain for odd $n_{\omega}$. Of course the $S=1$ chain may be transformed into a dimerized chain of $S=1/2$ blocks
by shifting one external carbon from one block to its right side neighbor block.
\section{Numerical studies}
We now turn to numerical verifications of that conjecture for
$n_{\omega}=1,2$ and 3 with several different techniques. 
Note that DMRG calculations have already been performed for the $n_{\omega}=1$ case~\cite{Ref5}. For larger 
$n_\omega$, the number of states to be kept becomes too large and the computation cost is prohibitive. Therefore, 
after providing quantitative estimates for the spin gap using approximate techniques, we have adressed this issues by using
an efficient QMC method. 
\subsection{Mapping into 1-D chains}
For $n_{\omega}=1$ the chain may be considered as built from $A$ and $B$ blocks of 7 and 9, 9 and 11, 11 and 13, or 13 and 15 sites, 
according to figure \ref{fig1} (B). One then obtains a dimerized $S=1/2$ spin chain with different values $J_1$, $J_2$ 
between 
the $A-B$ and $B-A$ blocks. The effective interactions $J_1$ and $J_2$ are directly obtained from the energy difference between the lowest 
triplet and singlet energies of $A-B$ and $B-A$ blocks~:
\begin{center}
\begin{tabular}{|c|c|c|c|c|}
\hline
 Blocks & 7-9 & 9-11 & 11-13 & 13-15 \\
\hline
$J_1$ & 0.221 & 0.196 & 0.181 & 0.170 \\
\hline
$J_2$ & 0.166 & 0.123 & 0.095 & 0.075 \\
\hline
\end{tabular}
\end{center}
A previons study of the dimerized $S=1/2$ chain has suggested that the gap follows the law~\cite{Ref7}~:
\begin{equation}
\Delta E=(J_1+J_2)\delta^{0.71} \ \mbox{with} \ \delta=\frac{(J_1-J_2)}{(J_1+J_2)}. 
\label{eq1}
\end{equation}
Applying Eq.~(\ref{eq1}) one obtains
\begin{eqnarray}
\Delta E & = & 0.097J \ \mbox{for the (7-9) blocks}, \nonumber \\
\Delta E & = & 0.112J \ \mbox{for the (9-11) blocks}, \nonumber \\
\Delta E & = & 0.121J \ \mbox{for the (11-13) blocks}, \nonumber\\
\Delta E & = & 0.126J \ \mbox{for the (13-15) blocks}, \nonumber
\end{eqnarray}
hence a finite gap.

For $n_{\omega}=2$ the simple mapping into a non-dimerized $S=1/2$ AF spin chain of 9-site blocks 
(cf. figure \ref{fig3} (B)) pleads in favor of a gapless character, but this
construction assumes only nearest-neighbor effective interaction $J_{NN}$, neglecting
for instance next-nearest neighbors interactions $J_{NNN}$. In fact, an uniform AF chain becomes gapped when the ratio $J_{NNN}/J_{NN}$ is larger 
than 0.241~\cite{Ref10}. We have extracted the effective couplings between 9-site blocks 
(figure \ref{fig3} (B)) from the exact 
spectrum of the trimer of blocks and we found an AF coupling between $NN$ blocks $J_{NN}=0.15896$ 
(in good agreement with the value extracted from the dimer $J_{NN}=0.16622$) and a surprisingly large 
ferromagnetic $J'_{NNN}=-0.1705$ coupling. Since an AF $S=1/2$ chain with 
ferromagnetic coupling between $NNN$ sites is not gapped, the $n_{\omega}=2$ ribbon should be gapless.

\begin{figure}[h]
\centerline{\includegraphics[scale=1.0]{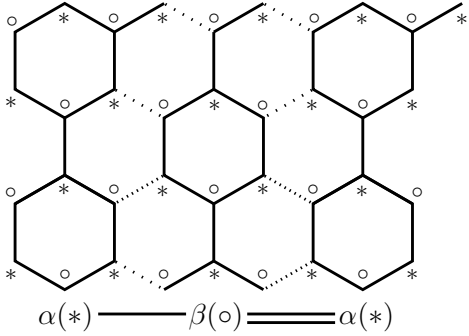}}
\caption{Definition of the blocks for a mapping into a dimerized $S=1/2$ chain of $n_{\omega}=3$ ribbon.}
\label{fig5}
\end{figure}

For $n_{\omega}=3$ one may define a dimerized AF $S=1/2$ chain of 13 and 11 sites respectively as pictured in 
figure \ref{fig5}. 
One obtains two values of the inter block AF coupling ($J_1=0.115$, $J_2=0.108$). Using equation~(\ref{eq1}) one obtains 
$\Delta E(n_{\omega}=3)=0.018J$.
\subsection{Renormalized excitonic calculations}
The recently proposed renormalized excitonic method \cite{Ref5bis} is again based on a periodic partition 
into blocks, but now the blocks have an even number of sites, a singlet ground state $\psi_A^0$ and a triplet 
lowest excited state $\psi_A^{\ast}$. One defines a model space for the $AB$ dimers made of two adjacent blocks, spanned by local singly excited 
states $\psi_A^{\ast}\psi_B^0$ and $\psi_A^0\psi_B^{\ast}$. Knowing the spectrum of the $AB$ dimer, it is possible to define
\cite{Ref11}
\begin{itemize}
\item[-] the effective energy of $\psi_A^{\ast}\psi_B^0$ and $\psi_A^0\psi_B^{\ast}$,
\item[-] the effective interaction between them.
\end{itemize}
These informations allow to apply the excitonic method. For the infinite lattice, this method leads to the following expressions 
of the excitation energy 
\begin{equation}
\Delta E^{\infty}(N_S)=2\Delta E(2N_S)-\Delta E(N_S).
\end{equation}
obtained from the gap values for a $N_S$-site block ($\Delta E(N_S)$) and for a dimer ($\Delta E(2 N_S)$).
\begin{figure}[t]
\centerline{\includegraphics[scale=1.0]{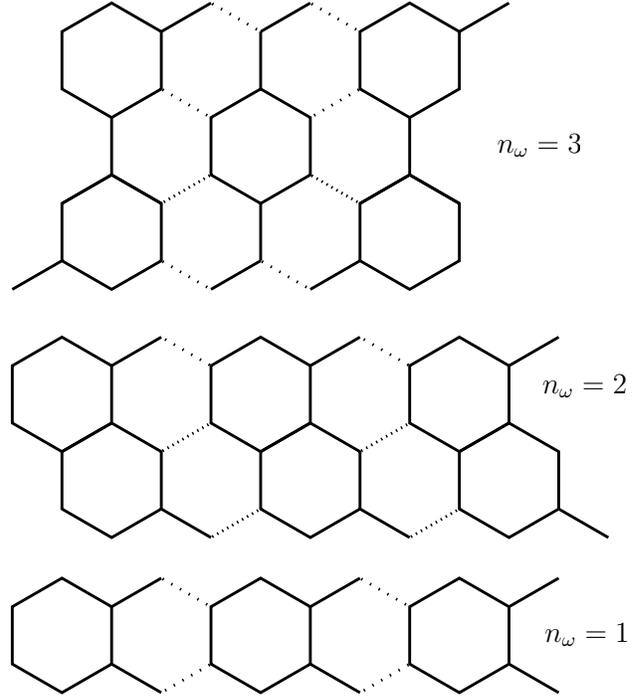}}
\caption{Definition of blocks for the calculation of the energy gap through the REM.}
\label{fig6}
\end{figure}

The method has been applied with success to the 2-leg ladders \cite{Ref8}
and can be applied to 
$n_{\omega}=1$, $n_{\omega}=2$ and $n_{\omega}=3$ polybenzeno\"id ribbons using the design of the blocks pictured in 
figure \ref{fig6}. 

We find a finite gap for $n_{\omega}=1$ polyacene:
\begin{eqnarray}
\Delta E^{\infty}(8) & = & 0.068J \ \mbox{for} \ N_S=8 \ \mbox{sites blocks,} \nonumber \\
\Delta E^{\infty}(12) & = & 0.094J \ \mbox{for} \ N_S=12 \ \mbox{sites blocks.} \nonumber 
\end{eqnarray}
Since the $N_S^{-1}$ components of the excitation energy cancels in the expression of $\Delta E^{\infty}(N_S)$, 
an extrapolation in terms of $N_S^{-2}$ leads to $\Delta E (n_\omega=1)=0.103J$. 

The result of the REM method for the $n_{\omega}=2$ lattice from the $N_S=12$ sites blocks 
is one order of magnitude smaller,
$\Delta E^{\infty}(12)=0.013J$. Extrapolation is not possible in 
this problem, but this result supports the conjecture that this fused polybenzeno\"id ribbon is not gapped.

For $n_{\omega}=3$ unequal blocks of $N_S=14$ and $N_S=10$ sites have been used, according to figure \ref{fig6}. 
Since the blocks 
are different it is necessary to generalize the algebra of Ref. 6, as performed in the Appendix. The final result is 
$\Delta E^{\infty}(14,10)=0.0125J$. Even if the results for the gaps for
$n_\omega=2$ and $3$ are quite close, these calculations support the idea of
a constrast between the odd and even ribbons. In order to confirm
this conjecture, we now present accurate Quantum Monte Carlo results for the $n_w=1,2$ and $3$
ribbons.

\subsection{Quantum Monte Carlo}
\begin{figure}[t]
\centerline{\includegraphics[width=0.48\textwidth]{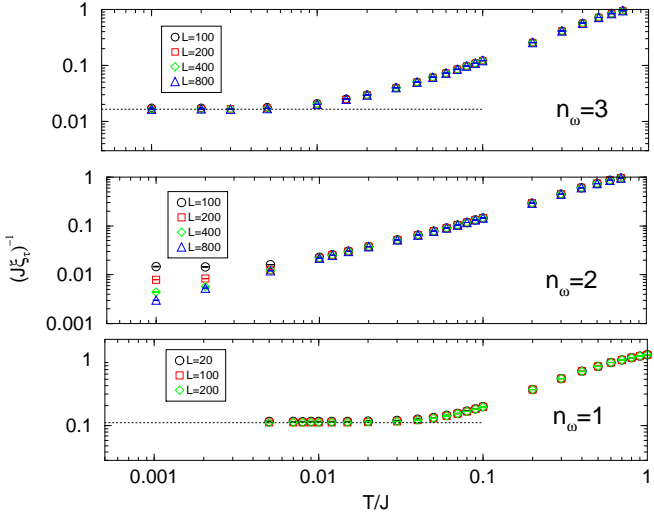}}
\caption{Quantum Monte Carlo results for the inverse of the imaginary time
  correlation length as a function of temperature for the $n_\omega=1$
  (bottom panel), $n_\omega=2$ (middle panel) and $n_\omega=3$ (top
  panel), for different system sizes $L$. Values of the extracted gaps
  $\Delta E (n_\omega=1)=0.1112(7) J$ and $\Delta E (n_\omega=3)=0.0164(4) J$ are represented by dashed-lines for $n_\omega=1$ and $3$.}
\label{fig:xit}
\end{figure}

As the spin lattices are not frustrated, efficient QMC algorithms
are available which allow to simulate with high-precision large systems at finite, albeit
extremely small, temperatures. Here we use a multi-cluster continuous
time \cite{beard} loop algorithm~\cite{evertz} which is free of any
systematic errors. We simulate systems of sizes up to length $L=800$ (the total
number of spins is $N=2(n_\omega+1)L$) and at inverse temperature up to
$\beta J=J/T=1000$. We use periodic boundary conditions along the legs. To determine the gapfull/gapless nature of the systems, we
calculate the correlation length in imaginary time $\xi_\tau$ with the help
of a standard second moment estimator~\cite{cooper,todo}
\begin{equation}
\xi_\tau = \frac{\beta}{2 \pi} \left(
\frac{\chi(\omega=0)}{\chi(\omega=2\pi/\beta)}-1 \right) ^{1/2}
\end{equation}
where $\chi(\omega)=\int_0^\beta d\tau e^{i\omega \tau} \chi(\tau)$ is the Fourier-transform
of the imaginary time dynamical antiferromagnetic structure factor
$\chi(\tau)=\frac{1}{\beta N}\sum_{i,j} (-)^{r_j-r_i} \int_O^\beta dt 
S^z_i(t) S^z_j(t+\tau)$. It can be shown that for a gapped system,
$\xi_\tau$ converges to the inverse spin gap in the thermodynamic limit at zero
temperature : $\lim_{L,\beta \rightarrow \infty} \xi_\tau(L,\beta) =
(\Delta E)^{-1}$. On the other hand, when the system is gapless, $\xi_\tau^{-1}$ is an
upper bound of the finite-size gap for any finite $L$ and $\beta$. In fig.~\ref{fig:xit}, we represent the inverse of the imaginary
correlation length $(J\xi_\tau)^{-1}$ as a function of the temperature $T/J$
in log-log scale, for the three types of ribbons $n_\omega=1,2$ and $3$. This representation is useful to see if the system is
gapped as $(J\xi_\tau)^{-1}$ saturates at low temperatures to the gap value
$\Delta E/J$. 

For $n_\omega=1$, $(J\xi_\tau)^{-1}$ clearly converges at low $T$ to a
minimum value identical for system sizes $L=100$ and $L=200$, indicating
that finite-size effects are absent. From the results for the largest $L$ at
the lowest $T$, we extract the value of the spin gap $\Delta E (n_\omega=1)=0.1112(7) J$, in perfect agreement with DMRG calculations~\cite{Ref5}.

We find no saturation of $(J\xi_\tau)^{-1}$ at the lowest temperature for
the largest system size $L=800$ studied for $n_\omega=2$. Data for smaller
systems ($L=100,200$) present signs of saturation at low enough $T$ towards
values which depend on the system size: this is naturally interpreted as the
signature of finite-size gaps. Strictly speaking, the numerical data for the
largest $L$ at the lowest $T=0.001$ can only put an upper bound
$\Delta E (n_\omega=2)<0.0030 J$ on the value of the gap. However, the general
form of the $T$ dependence of $(J\xi_\tau)^{-1}$ and the fact that we still
have finite-size effects for large systems at low $T$ naturally indicate
that the $n_{\omega}=2$ system is gapless, {\it i.e.} $\Delta E(n_\omega=2)=0$.

Finally for $n_\omega=3$, a convergence of $(J\xi_\tau)^{-1}$ is recovered
at low enough $T$ towards a size-independent constant. As for the
$n_{\omega}=1$ case, this indicates that the system is gapped and we obtain
the gap value $\Delta E (n_\omega=3)=0.0164(4) J$. Please note that we had
to resort to large systems at very low temperatures ($L=800$ and $T=0.001$)
to assert the complete convergence of our data: this is ascribed to the
small value of the gap (an order of magnitude lower than for
$n_{\omega}=1$). 

In conclusion, the QMC simulations unambiguously proove the even/odd number
$n_\omega$ effect for the gapless/gapfull nature of 1-D fused polyacenic
ribbons. As for spin ladders, we find that the value of the gap decreases
with the number $n_{\omega}$ of rings along the width for odd $n_{\omega}$ ribbons.
\section{Conclusion}
The conjecture of the existence of parity law concerning the spin gap in
polybenzeno\"id ribbons (vanishing / finite  gap for even / odd  $n_\omega$)
was proposed on the ground of qualitative arguments. This conjecture is confirmed by a consistent set of numerical calculations summarized in 
table 1.The REM fails to give a zero gap for
$n_{\omega}=2$ but indicates a contrast between odd and even ribbons. The quantitative mapping and the accurate QMC
calculations confirm the gapless character of the $n_{\omega}=2$ ribbons and agree on the amplitude of the gap for 
$n_{\omega}=1$ and $n_{\omega}=3$. For graphite ribbons~\cite{Ref2,Ref3}, for which $J\sim
2.2$ eV at the typical $r_{cc}$ distance (1.395\AA), 
the gaps should be close to $0.23$ eV for $n_{\omega}=1$ and $0.03$~eV for $n_{\omega}=3$.
Despite the semi-metallic nature of graphite, it seems unlikely that the
odd-$n_\omega$ ribbons present a zero charge gap, below their finite
spin-gap. We finally note, in analogy with the study of the influence
of four-spin operators in spin ladders~\cite{4spin}, that it would be worth considering
the possible impact of six-spin operators on such graphitic ribbons since
their amplitude is not negligible in hexagons~\cite{Maynau1,Maynau2}.

\begin{table}[h]
\begin{center}
\begin{tabular}{|c|ccc|}
\hline
 & Mapping & REM & QMC  \\
\hline
$n_{\omega}=1$ & 0.126 & 0.103 & 0.1112(7) \\
\hline
$n_{\omega}=2$ & 0 & 0.013 & 0 \\
\hline
$n_{\omega}=3$ & 0.018 & 0.0125 & 0.0164(4) \\
\hline
\end{tabular}
\end{center}
\caption{Calculed gaps (in units of $J$) obtained from different methods.}
\end{table}

\section{Acknowledgments}
The QMC calculations were performed using the looper application~\cite{todo}
(see http://wistaria.comp-phys.org/alps-looper ) of the ALPS
library~\cite{ALPS} (see http://alps.comp-phys.org). 
We warmly thank A. L\"auchli for suggesting the LSM argument for even $n_\omega$. 

\appendix
\section{Renormalized excitonic method for an $(A-B)_n$ chain}
One has two types of blocks $A$ and $B$. The ground state and lowest excited eigenfunctions for each block are given by
\begin{eqnarray*}
H_A \vert\psi^{0}_{A}\rangle & = & E^{0}_{A} \vert\psi^{0}_{A}\rangle,  \\
H_A\vert\psi_A^{\ast}\rangle & = & E^{\ast}_{A}\vert\psi_A^{\ast}\rangle,  \\
H_B \vert\psi^{0}_{B}\rangle & = & E^{0}_{B} \vert\psi^{0}_{B}\rangle,  \\
H_B\vert\psi_B^{\ast}\rangle & = & E^{\ast}_{B}\vert\psi_B^{\ast}\rangle.  
\end{eqnarray*} 
The ground state of the chain is represented by
\begin{equation*}
\Psi^0=\prod_i\psi^{0}_{A_i}\prod_j\psi^{0}_{B_j}. 
\end{equation*}
Its energy will be 
\begin{equation*}
\langle \Psi^0 \vert H^{eff} \vert \Psi^0 \rangle= n(E^{0}_{A}+E^{0}_{B})+n(v_{AB}+v_{BA}).
\end{equation*}
The interaction energies between two adjacent blocks is given by the knowledge of the exact energies of the 
$AB$ and $BA$ dimers
\begin{equation*} 
H_{AB}\vert \Psi^0_{AB} \rangle = E^{0}_{AB}\vert \Psi^0_{AB} \rangle, \ E^{0}_{AB}=E^{0}_{A}+E^{0}_{B}+v_{AB}.
\end{equation*}
For the description of excited states one needs to estimate the effective interaction between a local excited 
state and the neighbor ground states and the integral responsible for the transfer of excitation. These 
informations are obtained from the excited solutions of the dimers. One should especially consider the two eigenstates 
\begin{eqnarray*}
H_{AB}\vert\Psi_{AB}^{\ast}\rangle & = & E^{\ast}_{AB}\vert\Psi_{AB}^{\ast}\rangle, \\ 
H_{AB}\vert\Psi_{AB}^{\ast'}\rangle & = & E^{\ast'}_{AB}\vert\Psi_{AB}^{\ast'}\rangle,
\end{eqnarray*}
having the largest projections $\vert \tilde{\Psi}_{AB}^{\ast} \rangle$ and $\vert \tilde{\Psi}_{AB}^{\ast'} \rangle$, 
on the model space spanned by $\psi_{A}^{\ast}\psi_{B}^{0}$ and 
$\psi_{A}^{0}\psi_{B}^{\ast}$, which can be written after orthogonalization as 
\begin{eqnarray*}
\vert \tilde{\Psi}_{AB}^{\ast} \rangle & = & \lambda \vert \psi_A^{\ast} \psi_B^{0}\rangle + \mu 
\vert \psi_A^{0} \psi_B^{\ast}\rangle,  \\
\vert \tilde{\Psi}_{AB}^{\ast'} \rangle & = & -\mu \vert \psi_A^{\ast} \psi_B^{0}\rangle + \lambda
\vert \psi_A^{0} \psi_B^{\ast}\rangle.  
\end{eqnarray*}
It results that
\begin{eqnarray*}
\langle \psi_{A}^{\ast}\psi_B^{0} \vert H^{eff}\vert \psi_{A}^{\ast}\psi_B^{0} \rangle & = &
\lambda^2 E^{\ast}_{AB}+\mu^2 E^{\ast'}_{AB} \nonumber \\
& = & E^{\ast}_{A}+E^{0}_{B}+v_{(A^{\ast})B},   \\
\langle \psi_{A}^{0}\psi_B^{\ast} \vert H^{eff}\vert \psi_{A}^{0}\psi_B^{\ast} \rangle & = &
\mu^2 E^{\ast}_{AB}+\lambda^2 E^{\ast'}_{AB} \nonumber \\
& = & E^{0}_{A}+E^{\ast}_{B}+v_{A(B^{\ast})},  \\
\langle \psi_{A}^{\ast}\psi_B^{0} \vert H^{eff}\vert \psi_{A}^{0}\psi_B^{\ast} \rangle & = &
(E^{\ast}_{AB}-E^{\ast'}_{AB})\lambda\mu = h_{AB}. 
\end{eqnarray*}
For the periodic system the delocalized excited states will be respresented as linear combinations 
of locally excited states on either $A$ or $B$ blocks
\begin{eqnarray*}
\Psi^{\ast}_{A_m} & = & \psi^{\ast}_{A_m}\prod_{i\neq m}\psi^{0}_{A_i}\prod_j\psi^{0}_{B_j}, \\
\Psi^{\ast}_{B_n} & = & \psi^{\ast}_{B_n}\prod_{j\neq n}\psi^{0}_{B_j}\prod_i\psi^{0}_{A_i}, 
\end{eqnarray*}
The energies of $\Psi^{\ast}_{A_m}$ and $\Psi^{\ast}_{B_n}$ are given by
\begin{eqnarray*}
\langle \Psi^{\ast}_{A_m} \vert H^{eff} \vert \Psi^{\ast}_{A_m} \rangle -
\langle \Psi^0 \vert H^{eff} \vert \Psi^0 \rangle  = \nonumber \\
E^{\ast}_{A}-E^{0}_{A}+v_{(A^{\ast})B}-v_{AB},  \\
\langle \Psi^{\ast}_{B_n} \vert H^{eff} \vert \Psi^{\ast}_{B_n} \rangle -
\langle \Psi^0 \vert H^{eff} \vert \Psi^0 \rangle =  \nonumber \\
 E^{\ast}_{B}-E^{0}_{B}+v_{(B^{\ast})A}-v_{BA},  
\end{eqnarray*}
This locally excited state are coupled with the states locally excited on the adjacent $B$ blocks
\begin{equation*}
\langle \Psi^{\ast}_{A_m} \vert H^{eff} \vert \Psi^{\ast}_{B_m} \rangle = h_{AB}.
\end{equation*}
For the lowest state of the lattice, corresponding to $\overrightarrow{k}=0$, the delocalized excited states can be 
written as a linear combination of 
\begin{equation*}
(\Psi^{\ast}_a)_{\overrightarrow{k}=0}  = \frac{1}{\sqrt N}\sum_{A_m}\Psi^{\ast}_{A_m}, 
\end{equation*}
and
\begin{equation*}
(\Psi^{\ast}_b)_{\overrightarrow{k}=0}  =  \frac{1}{\sqrt N}\sum_{B_n}\Psi^{\ast}_{B_n},
\end{equation*}
solution of a $2\times 2$ matrix whose elements are 
\begin{eqnarray*}
\Big\langle (\Psi^{\ast}_a)_{\overrightarrow{k}=0} \Big\vert H^{eff} \Big\vert (\Psi^{\ast}_a)_{\overrightarrow{k}=0} 
\Big\rangle - \Big\langle \Psi^0 \Big\vert H^{eff} \Big\vert \Psi^0 \Big\rangle  = \nonumber \\ 
E^{\ast}_{A}-E^{0}_{A}+2\Big(V_{(A^{\ast})B}-V_{AB}\Big),  \\
\Big\langle (\Psi^{\ast}_b)_{\overrightarrow{k}=0} \Big\vert H^{eff} \Big\vert (\Psi^{\ast}_b)_{\overrightarrow{k}=0} 
\Big\rangle - \Big\langle \Psi^0 \Big\vert H^{eff} \Big\vert \Psi^0 \Big\rangle  =  \nonumber \\
E^{\ast}_{B}-E^{0}_{B}+2\Big(V_{(B^{\ast})A}-V_{BA}\Big),  \\
\Big\langle (\Psi^{\ast}_a)_{\overrightarrow{k}=0} \Big\vert H^{eff} \Big\vert (\Psi^{\ast}_b)_{\overrightarrow{k}=0} 
\Big\rangle  =  2h_{AB}.
\end{eqnarray*}

\end{document}